\input psfig.sty
\documentstyle{cupconf}

\ifoldfss
\else
  \ifnfssone
    \newmathalphabet{\mathit}
      \addtoversion{normal}{\mathit}{cmr}{m}{it}
      \addtoversion{bold}{\mathit}{cmr}{bx}{it}
    \newmathalphabet{\mathcal}
      \addtoversion{normal}{\mathcal}{cmsy}{m}{n}
    \else
    \ifnfsstwo
    \fi
  \fi
\fi

\def\HI{H{\,\small I}}
\def\slHI{H{\,\small\sl I}}

\def\kms{km s$^{-1}$}

\def\hexnumber#1{\ifcase#1 0\or1\or2\or3\or4\or5\or6\or7\or8\or9\or
 A\or B\or C\or D\or E\or F\fi }

\makeatletter
\ifx\CUP@mtlplain@loaded\undefined
\else
\fi
\makeatother
 \makeatletter
 \ifx\CUP@mtlplain@loaded\undefined
   \font\tenbmi=cmmib10 at 10pt
   \font\sevenbmi=cmmib10 at 7pt
   \font\fivebmi=cmmib10 at 5pt

   \newfam\bmifam
   \textfont\bmifam=\tenbmi
   \scriptfont\bmifam=\sevenbmi
   \scriptscriptfont\bmifam=\fivebmi
   
 \fi
 \makeatother
\ifnfsstwo

\fi
\ifnfssone

\fi
\ifoldfss

\fi

\mathchardef\varLambda="0103

\makeatletter
\ifx\CUP@mtlplain@loaded\undefined
\else
\fi
\makeatother
\makeatletter
\ifx\CUP@mtlplain@loaded\undefined
  \font\tenbms=cmbsy10
  \font\sevenbms=cmbsy10 at 7pt
  \font\fivebms=cmbsy10 at 5pt
  \newfam\bmsfam
  \textfont\bmsfam=\tenbms
  \scriptfont\bmsfam=\sevenbms
  \scriptscriptfont\bmsfam=\fivebms

  \edef\bsy@{\hexnumber\bmsfam}
  \mathchardef\bnabla="0\bsy@72
\fi
\makeatother
\title[HI absorption in radio galaxies]
{HI absorption and the ISM around radio galaxies}

\author[Morganti et al.]{%
R. Morganti$^1$, T. Oosterloo$^1$, C.N. Tadhunter$^2$, K.A. Wills$^2$, 
A.Tzioumis$^3$, J. Reynolds$^3$}

\affiliation{$^1$ ASTRON - NFRA, Postbus 2, 7990 AA
Dwingeloo, The Netherlands \\
$^2$ Dep. Physics and Astronomy,
University of Sheffield,  S7 3RH, UK \\
$^3$ Australia Telescope National Facility, CSIRO, P.O. Box 76, 2121 Epping
NSW, Australia}

\begin{document}
\ifnfssone
\else
  \ifnfsstwo
  \else
    \ifoldfss
      \let\mathcal\cal
      \let\mathrm\rm
      \let\mathsf\sf
    \fi
  \fi
\fi

\maketitle

\begin{abstract}
  
  We present a study, done with the Australian LBA, of  \HI\ absorption  for
  two compact radio galaxies (PKS~1549-79 and PKS~1814-63).  In both the radio
  galaxies, the \HI\ appears to give us information about the
  environment in which the radio sources are embedded, the effect that the ISM
  can have on the observed characteristics and the possible presence of
  interaction between the ISM and the radio plasma.

\end{abstract}

\firstsection 
\section{Introduction}

\HI\ absorption is often detected in compact/small radio galaxies.  Indeed,
the scale of the radio continuum in these objects represents an effective
background against which the absorption can be observed.  Although the \HI\
absorption is often interpreted as due to the presence of a nuclear
torus/disk, the situation can be more complicated in some (or perhaps many)
cases.  For example, extended regions of ionised gas have been imaged around
compact radio galaxies (e.g.  with HST, see Axon et al.\ 2000).  If this gas
is also a tracer of the neutral gas, this means that \HI\ absorption can also
be produced in the regions around the radio lobes and not only against the
nucleus.  Moreover, compact radio galaxies (e.g.\ Compact Steep Spectrum
sources) are known to have broad (forbidden) optical emission lines likely due
to turbulent motions as result of the interaction between the radio plasma and
the ISM (Gelderman \& Whittle 1994).  This process may ``pile up'' gas at the
edge of the lobes, thus making the detection of \HI\ absorption more likely
(see e.g.\ IC~5063, Oosterloo et al.\ 2000).  In some cases (although a
minority) star formation and far-IR emission are observed.  Thus, these
objects could have a particularly rich ISM that could produce the \HI\
absorption in/around their radio lobes. 

A further complication in the interpretation of the \HI\ absorption is the
uncertainty in the values of the systemic velocity as obtained from optical
data.  To unambiguously interpret the kinematics of the neutral gas, better
systemic velocities than available in literature are usually necessary. 

We have recently carried out (with the VLA and the ATCA) a survey looking for
HI absorption in radio galaxies: 23 galaxies have been observed so far
(Morganti et al.\ 2000a).  Two of them are compact radio galaxies and show
\HI\ absorption (PKS 1549--79 and PKS 1814--63).  Thus, they have been
followed up with high resolution observations with the Australian LBA. 

\section{PKS~1549--79}

PKS~1549--79 is a relatively unusual galaxy.  At radio wavelengths, it is a
compact flat spectrum source ($\alpha =0.35$, $S \sim \nu ^{-\alpha }$) with a
one-sided jet on the VLBI scale.  The radio
structure seems to indicate that the radio jet axis is aligned with our line
of sight.  The size is 150 mas, about 350 pc \footnote{In this paper we assume
$H_\circ = 50$ \kms Mpc$^{-1}$ and $q_\circ = 0$.}.  Because of these
characteristics, it was quite surprising to detect \HI\ absorption in this
object. 

PKS 1549--79 has a redshift of $z=0.150$ and was therefore observed at
1235~MHz, the frequency of the redshifted \HI.  At this relatively low
frequency we could use only three telescopes Parkes (64m), CA (Tied Array,
5x22m) and Mopra (22m).  The latter had technical problems so data on only one
baseline was left.  The observations were done using 8~MHz band, 512 channels
($\sim 3$ \kms\ velocity resolution). 
The \HI\ profile is shown in Fig.1. The total of the line is $\sim 200$ \kms\
and the peak optical depth is about 2\%.  The column density is
$\sim 4\times 10^{18} T_{\rm spin}$ cm$^{-2}$.

\begin{figure}[!h]
\centerline{\psfig{figure=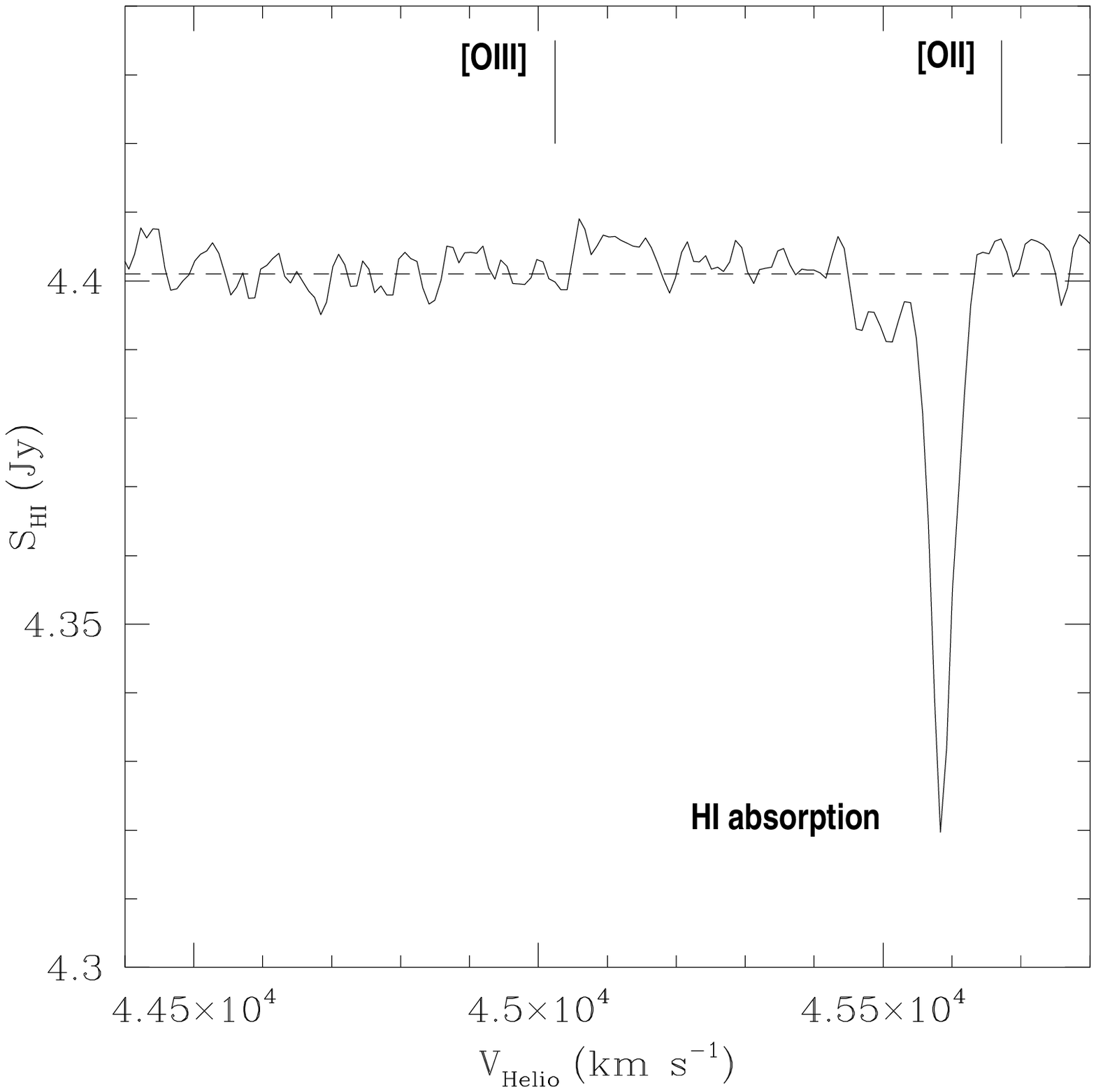,width=4.5cm}
\psfig{figure=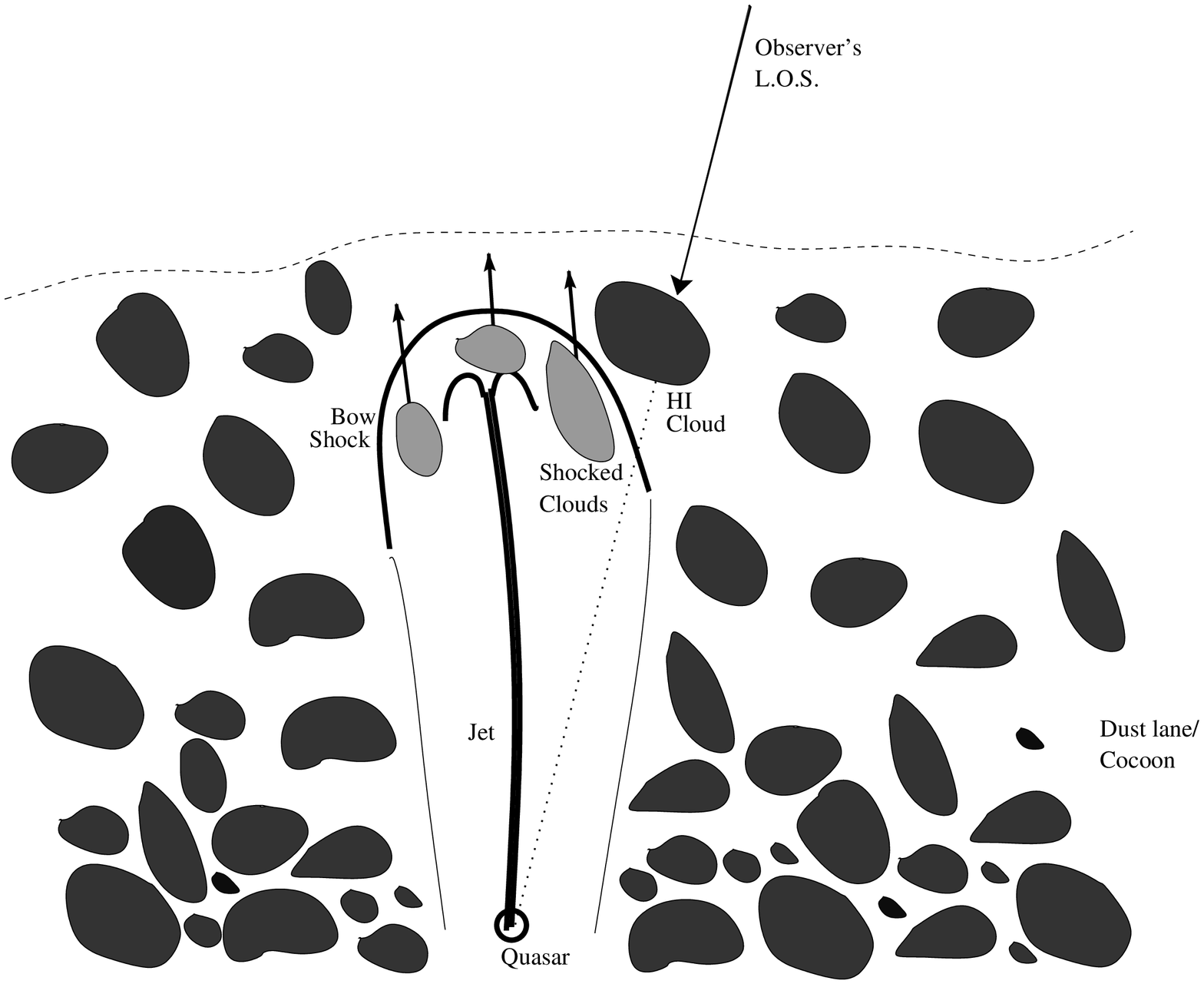,width=7cm,angle=-90}
}
 \caption{{\it Left:} profile of the \HI\ absorption in PKS~1549--79. The velocities
derived from the optical emission lines are indicated. {\it Right:} cartoon of
the possible distribution of the various components in PKS~1549--79, see text
for details. }
 \end{figure}

A high quality optical spectrum was taken (see Tadhunter et al.\ 2000a for
details) to allow an accurate comparison between the velocities derived from
the optical lines (ionised gas) and from the \HI.  The spectrum of PKS~1549--79
is characterised by high ionisation ``narrow'' emission lines with {\sl no
evidence for broad permitted lines} (quite unusual for a flat spectrum radio
source).  Surprisingly, {\sl two redshift systems} were found in the optical:
the higher ionisation lines (e.g.\ [OIII]5007\AA) have a significant lower
redshift (velocity difference of $\Delta v = 600$ \kms, see Fig.\ 1) than the
low ionisation lines (e.g.\ [OII]3727\AA). The {\sl high ionisation forbidden
line} are also unusually broad (1345 \kms). {\sl Interestingly, the velocity
derived from the low ionisation lines is consistent with the one derived from
the \slHI}. PKS~1549--79 is also one of the few known radio galaxies at low
redshift which shows (in the optical continuum) a strong component from young
stellar population. This is confirmed also by the strong far-IR emission
detected by the IRAS satellite (Roy \& Norris 1997). 

The likely scenario of what is going on in PKS~1549--79, in order to explain
{\sl all} the observed characteristics, is summarised in the cartoon in Fig.\ 
1.  PKS~1549--79 is a {\sl young} source where the nucleus is surrounded by a
cocoon of material left over from the events which triggered the nuclear
activity.  The high ionisation lines are formed in a region close to the
central AGN, which is undergoing outflow because of interactions, e.g., with
the radio jet, while the low ionisation lines are formed in a obscuring region
at larger distance and not so disturbed kinematically.  As the radio source
evolves, any obscuring material along the jet axis is likely to be swept aside
by, e.g., jet-cloud interactions.  Before this stage is reached a substantial
amount of obscuring material may be present along the radio axis.  A major
implication of this work is that the simplest version of the unified schemes,
in which lines of sight close to the radio axis have a relatively unobscured
view of the quasar nucleus, {\sl may not always hold for young, compact radio
sources in which the jets and quasar winds have not yet swept aside the warm
ISM in the ionisation cones}.

Although the characteristics observed in PKS~1549--79 may not be very common
among radio galaxies, we have found at least an other object with remarkable
similarities: the compact radio source 4C~12.30 (PKS~1345+12).

\section{PKS~1814--63}

PKS~1814--63 is a Compact Steep Spectrum (CSS) source with a basic
double-lobed structure oriented almost north-south (Tzioumis et al.\  1996).
The radio spectral index on the VLBI scale is not yet available for this
galaxy, thus, no information is available on which of these structures (if
any) corresponds to the core.  The overall extent of the source is 410 mas
corresponding to 328 pc.  With ATCA  \HI\ absorption was detected that
shows a deep profile ($\sim$20 \% optical depth) together with a broad,
shallow feature (Morganti et al.\ 2000a).
PKS~1814--63 has a redshift of $z=0.06$ and was observed with the Australian
LBA at 1336~MHz using Parkes, CA (Tied Array), Mopra, Hobart with the
same setup as for PKS~1549--79. 

\begin{figure}[!h]
\centerline{\psfig{figure=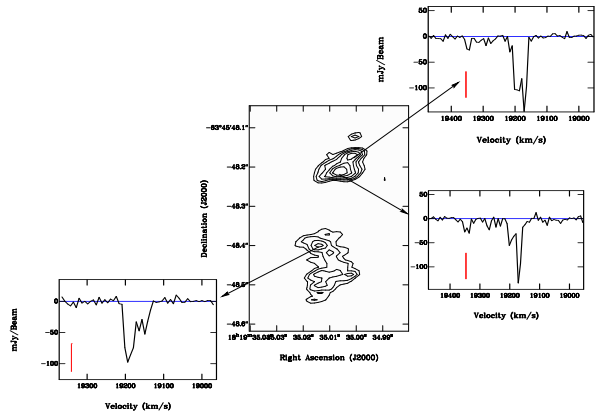,width=8cm}
}
 \caption{Continuum image of PKS~1814--63 at 21 cm (obtained from the
line-free channels) and spectra of the \HI\ absorption obtained at three
different location in the source.  The vertical line indicates the systemic velocity
(see text for details).}
 \end{figure}

The results are shown in Fig.\ 2.  The deep absorption appears extended in the
VLBI data and is detected against most of the source with the exception of the
southern part of the southern lobe.  The peak optical depth goes from $\sim
30$\% in the southern lobe to $\sim$ 10\% in the northern region.  The
component with lower optical depth (ranging between 2 and 4\%) is observed
{\sl only} against the northern region.
The optical spectra available for this object are not very good.  The systemic
velocity marked in the plots ($V_{\rm hel}=19350$ \kms) is derived from
re-analysing the optical spectrum presented in Tadhunter et al.\ (1993).  This
velocity will have to be confirmed by better quality data.  However, as it
stands now, most of the \HI\ absorption, and in particular the deep component,
is blueshifted compared to the systemic velocity.  Only the component with low
optical depth observed against the northern region has a velocity close to the
systemic velocity as derived from the available data. 

Thus, one possibility to explain these
characteristics is that the shallow component is the one associated with a
circumnuclear disk while the deeper component (observed against most of the
radio source) could be instead associated with more extended gas, possibly
surrounding the lobes and perhaps interacting/expanding with them (if the
blueshifted velocity is confirmed).

\section{\HI\ absorption and ISM medium}

In both the radio galaxies studied, the \HI\ appears to give us information
about the environment in which the radio sources are embedded, the effect that
the ISM can have on the observed characteristics and the possible presence of
interaction between the ISM and the radio plasma.

\HI\ absorption occurring around radio lobes and/or related to interaction
between the radio plasma and the ISM is indeed found in a growing number of
radio sources.  An extreme case of interaction has been found in the Seyfert
galaxy IC~5063 where very broad ($\sim$700 \kms) \HI\ absorption, blueshifted
with respect to the systemic velocity has been observed coincident with a
radio lobe (Oosterloo et al.\ 2000).  Other possible evidence for interaction
between the radio plasma and the ISM has been found in at least two radio
galaxies: the superluminal object 3C 216 (Pihlstr\"om et al.\ 1999) and 3C 326
(Conway et al.\ these proceedings).  Moreover, on even larger scale ($>$ 10
kpc), \HI\ absorption against the radio lobes has been found in the radio
galaxy Coma~A (Morganti et al.\ 2000b) as well as in the radio galaxy 3C~433. 
In Coma~A, the striking structure of the ionised gas strongly suggests the
presence of a complex interaction between the radio structure and the gas
around (Tadhunter et al.  2000b) while 3C~433 is an other example of starburst
and far-IR bright radio galaxies).  If the \HI\ absorption is observed against
the lobes of relatively large galaxies situated in a relatively rich
environment, this situation could be even more common for compact radio
sources.  Thus, the study of the \HI\ absorption can tell us about the role of
the ISM in the evolution of the sources as well as help in understanding the
relation between intrinsic distribution of the gas and ionised gas.  Moreover,
it can give us clues on the ionisation mechanisms. 

This study also shows the importance of having information on as many as
possible wavebands and also good enough optical data for a proper
interpretation of the \HI\ data.

\end{document}